\documentclass[10pt,a4paper,onecolumn]{article}
\usepackage{marginnote}
\usepackage{graphicx}
\usepackage{xcolor}
\usepackage{authblk,etoolbox}
\usepackage{titlesec}
\usepackage{calc}
\usepackage{tikz}
\usepackage{hyperref}
\hypersetup{colorlinks,breaklinks=true,
            urlcolor=[rgb]{0.0, 0.5, 1.0},
            linkcolor=[rgb]{0.0, 0.5, 1.0},
            citecolor = blue}
\usepackage{caption}
\usepackage{tcolorbox}
\usepackage{amssymb,amsmath}
\usepackage{ifxetex,ifluatex}
\usepackage{seqsplit}
\usepackage{xstring}
\usepackage{natbib}
\usepackage[T1]{fontenc}
\usepackage[utf8]{inputenc}
\usepackage{cleveref}
\usepackage{float}
\usepackage{orcidlink}
\usepackage{xspace}
\usepackage{listings}
\usepackage{comment}
\usepackage{markdown}
\usepackage{tikz}
\usetikzlibrary{shapes,arrows,shadows,fit}
\usetikzlibrary{positioning}
\usetikzlibrary{bayesnet}

\let\origfigure\figure
\let\endorigfigure\endfigure
\renewenvironment{figure}[1][2] {
    \expandafter\origfigure\expandafter[H]
} {
    \endorigfigure
}

\let\textttOrig=\texttt
\def\texttt#1{\expandafter\textttOrig{\seqsplit{#1}}}
\renewcommand{\seqinsert}{\ifmmode
  \allowbreak
  \else\penalty6000\hspace{0pt plus 0.02em}\fi}

\usepackage{abstract}

\makeatletter
\let\href@Orig=\href
\def\href@Urllike#1#2{\href@Orig{#1}{\begingroup
    \def\Url@String{#2}\Url@FormatString
    \endgroup}}
\def\href@Notdoi#1#2{\def\tempa{#1}\def\tempb{#2}%
  \ifx\tempa\tempb\relax\href@Urllike{#1}{#2}\else
  \href@Orig{#1}{#2}\fi}
\def\href#1#2{%
  \IfBeginWith{#1}{https://doi.org}%
  {\href@Urllike{#1}{#2}}{\href@Notdoi{#1}{#2}}}
\makeatother

\newlength{\cslhangindent}
\newlength{\csllabelwidth}
 {
  \setlength{\parindent}{0pt}
  \ifodd #1 \everypar{\setlength{\hangindent}{\cslhangindent}}\ignorespaces\fi
  \ifnum #2 > 0
  \setlength{\parskip}{#2\baselineskip}
  \fi
 }%
 {}
\usepackage{calc}

\usepackage[top=3.5cm, bottom=3cm, right=1.5cm, left=1.0cm,
            headheight=2.2cm, reversemp, includemp, marginparwidth=4.5cm]{geometry}

\titleformat{\section}
  {\normalfont\sffamily\Large\bfseries}
  {}{0pt}{}
\titleformat{\subsection}
  {\normalfont\sffamily\large\bfseries}
  {}{0pt}{}
\titleformat{\subsubsection}
  {\normalfont\sffamily\bfseries}
  {}{0pt}{}
\titleformat*{\paragraph}
  {\sffamily\normalsize}

\usepackage{fancyhdr}
\makeatletter
\let\ps@plain\ps@fancy
\fancyheadoffset[L]{4.5cm}
\fancyfootoffset[L]{4.5cm}

\definecolor{linky}{rgb}{0.0, 0.5, 1.0}

\newtcolorbox{repobox}
   {colback=red, colframe=red!75!black,
     boxrule=0.5pt, arc=2pt, left=6pt, right=6pt, top=3pt, bottom=3pt}

\newcommand{\ExternalLink}{%
   \tikz[x=1.2ex, y=1.2ex, baseline=-0.05ex]{%
       \begin{scope}[x=1ex, y=1ex]
           \clip (-0.1,-0.1)
               --++ (-0, 1.2)
               --++ (0.6, 0)
               --++ (0, -0.6)
               --++ (0.6, 0)
               --++ (0, -1);
           \path[draw,
               line width = 0.5,
               rounded corners=0.5]
               (0,0) rectangle (1,1);
       \end{scope}
       \path[draw, line width = 0.5] (0.5, 0.5)
           -- (1, 1);
       \path[draw, line width = 0.5] (0.6, 1)
           -- (1, 1) -- (1, 0.6);
       }
   }

\patchcmd{\@maketitle}{center}{flushleft}{}{}
\patchcmd{\@maketitle}{center}{flushleft}{}{}
\patchcmd{\@maketitle}{\LARGE}{\LARGE\sffamily}{}{}

\def\maketitle{{%
  
  \g@addto@macro\AB@authors{%
    \protect\\[0.5em]%
    \protect\normalfont\protect\itshape on behalf of the Euclid Consortium%
  }%
  \AB@maketitle}}
  
\renewcommand\AB@affilsepx{ \protect\Affilfont}
\renewcommand\AB@affilnote[1]{{\bfseries #1}\hspace{3pt}}
\renewcommand{\affil}[2][]%
   {\newaffiltrue\let\AB@blk@and\AB@pand
      \if\relax#1\relax\def\AB@note{\AB@thenote}\else\def\AB@note{#1}%
        \setcounter{Maxaffil}{0}\fi
        \begingroup
        \let\href=\href@Orig
        \let\texttt=\textttOrig
        \let\protect\@unexpandable@protect
        \def\thanks{\protect\thanks}\def\footnote{\protect\footnote}%
        \@temptokena=\expandafter{\AB@authors}%
        {\def\\{\protect\\\protect\Affilfont}\xdef\AB@temp{#2}}%
         \xdef\AB@authors{\the\@temptokena\AB@las\AB@au@str
         \protect\\[\affilsep]\protect\Affilfont\AB@temp}%
         \gdef\AB@las{}\gdef\AB@au@str{}%
        {\def\\{, \ignorespaces}\xdef\AB@temp{#2}}%
        \@temptokena=\expandafter{\AB@affillist}%
        \xdef\AB@affillist{\the\@temptokena \AB@affilsep
          \AB@affilnote{\AB@note}\protect\Affilfont\AB@temp}%
      \endgroup
       \let\AB@affilsep\AB@affilsepx
}
\makeatother

\renewcommand\Affilfont{\sffamily\small\mdseries}
\ifnum 0\ifxetex 1\fi\ifluatex 1\fi=0 
  \usepackage[T1]{fontenc}
  \usepackage[utf8]{inputenc}

\else 
  \ifxetex
    \usepackage{mathspec}
    \usepackage{fontspec}

  \else
    \usepackage{fontspec}
  \fi
  \defaultfontfeatures{Ligatures=TeX,Scale=MatchLowercase}

\fi
\IfFileExists{upquote.sty}{\usepackage{upquote}}{}
\IfFileExists{microtype.sty}{%
\usepackage{microtype}
\UseMicrotypeSet[protrusion]{basicmath} 
}{}

\let\addcontentslineOrig=\addcontentsline
\def\addcontentsline#1#2#3{\bgroup
  \let\texttt=\textttOrig\addcontentslineOrig{#1}{#2}{#3}\egroup}
\let\markbothOrig\markboth
\def\markboth#1#2{\bgroup
  \let\texttt=\textttOrig\markbothOrig{#1}{#2}\egroup}
\let\markrightOrig\markright
\def\markright#1{\bgroup
  \let\texttt=\textttOrig\markrightOrig{#1}\egroup}

\usepackage{graphicx,grffile}
\makeatletter
\def\maxwidth{\ifdim\Gin@nat@width>\linewidth\linewidth\else\Gin@nat@width\fi}
\def\maxheight{\ifdim\Gin@nat@height>\textheight\textheight\else\Gin@nat@height\fi}
\makeatother
\setkeys{Gin}{width=\maxwidth,height=\maxheight,keepaspectratio}
\IfFileExists{parskip.sty}{%
\usepackage{parskip}
}{
\setlength{\parindent}{0pt}
\setlength{\parskip}{6pt plus 2pt minus 1pt}
}
\providecommand{\tightlist}{%
  \setlength{\itemsep}{0pt}\setlength{\parskip}{0pt}}
\ifx\paragraph\undefined\else
\let\oldparagraph\paragraph
\renewcommand{\paragraph}[1]{\oldparagraph{#1}\mbox{}}
\fi
\ifx\subparagraph\undefined\else
\let\oldsubparagraph\subparagraph
\renewcommand{\subparagraph}[1]{\oldsubparagraph{#1}\mbox{}}
\fi

\DeclareFixedFont{\ttb}{T1}{txtt}{bx}{n}{10} 
\DeclareFixedFont{\ttm}{T1}{txtt}{m}{n}{10}  

\usepackage{color}
\definecolor{deepblue}{rgb}{0,0,0.5}
\definecolor{deepred}{rgb}{0.6,0,0}
\definecolor{deepgreen}{rgb}{0,0.5,0}

\usepackage{listings}

\newcommand\pythonstyle{\lstset{
language=Python,
basicstyle=\scriptsize,
morekeywords={self},              
keywordstyle=\scriptsize\color{deepblue},
emph={MyClass,__init__},          
emphstyle=\ttb\color{deepred},    
stringstyle=\color{deepgreen},
    numbers=left,
frame=tb,                         
showstringspaces=false
}}

\lstnewenvironment{python}[1][]
{
\pythonstyle
\lstset{#1}
}
{}

\newcommand\pythoninline[1]{{\pythonstyle\lstinline!#1!}}

\newcommand{\pkgfont}{\texttt} 
\newcommand{\pkg}[2][]{%
  \if\relax\detokenize{#1}\relax
    \pkgfont{#2}%
  \else
    \href{#1}{\pkgfont{#2}}%
  \fi
}

\usepackage{listings}
\usepackage{xcolor}

\lstdefinelanguage{YAML}{
  keywords={true, false, null},
  keywordstyle=\color{blue}\bfseries,
  comment=[l]{\#},
  commentstyle=\color{gray}\itshape,
  stringstyle=\color{teal},
  morestring=[b]",
  morestring=[b]',
}

\newcommand*{\AckInstitutions}{a number of agencies and
  institutes that have supported the development of \textit{Euclid}, in
  particular
  the Agenzia Spaziale Italiana,
  the Austrian Forschungsf\"orderungsgesellschaft funded through BMIMI,
  the Belgian Science Policy,
  the Canadian Euclid Consortium,
  the Deutsches Zentrum f\"ur Luft- und Raumfahrt,
  the DTU Space and the Niels Bohr Institute in Denmark,
  the French Centre National d'Etudes Spatiales,
  the Funda\c{c}\~{a}o para a Ci\^{e}ncia e a Tecnologia,
  the Hungarian Academy of Sciences,
  the Ministerio de Ciencia, Innovaci\'{o}n y Universidades,
  the National Aeronautics and Space Administration,
  the National Astronomical Observatory of Japan,
  the Netherlandse Onderzoekschool Voor Astronomie,
  the Norwegian Space Agency,
  the Research Council of Finland,
  the Romanian Space Agency,
  the Swiss Space Office (SSO) at the State Secretariat for Education, Research, and Innovation (SERI),
  and the United Kingdom Space Agency.
  A complete and detailed list is available on the \textit{Euclid}\ web site
  (\url{www.euclid-ec.org/consortium/community/}).\xspace}
  
\newcommand{\AckEC}{The Euclid Consortium acknowledges the European
  Space Agency and \AckInstitutions}

\title{\texttt{cloelike}: A Python Library for Cosmological Likelihood Inference in the Euclid Era}

\date{\vspace{-7ex}}

\begin{document}
\author[1]{(cloe-org maintainers:) Marco Bonici\orcidlink{0000-0002-8430-126X}}
\author[2]{Guadalupe Cañas-Herrera\thanks{Corresponding author: canasherrera@strw.leidenuniv.nl}\orcidlink{0000-0003-2796-2149}}
\author[3]{Pedro Carrilho\orcidlink{0000-0003-1339-0194}}
\author[4]{Santiago Casas\orcidlink{0000-0002-4751-5138}}
\author[5]{Chiara Moretti\orcidlink{0000-0003-3314-8936}}
\author[6]{Andrea Pezzotta\orcidlink{0000-0003-0726-2268}}

\author[7]{(cloe-org contributors:) Zahra Baghkhani\orcidlink{0000-0002-6632-2614}}
\author[8]{Carmelita Carbone\orcidlink{0000-0003-0125-3563}}
\author[7]{Martin Crocce\orcidlink{0000-0002-9745-6228}}
\author[2]{Jip de Buck\orcidlink{0009-0001-5175-9282}}
\author[9]{Klara Bertmann\orcidlink{0009-0004-6700-2470}}
\author[10]{Nastassia Grimm\orcidlink{0000-0001-9602-0599}}
\author[11]{Martin Kärcher\orcidlink{0000-0001-5868-647X}}
\author[12]{Felicitas Keil\orcidlink{0000-0002-8108-1679}}
\author[13]{Davide Sciotti\orcidlink{0009-0008-4519-2620}}
\author[14]{Peter L. Taylor\orcidlink{0000-0001-6999-4718}}
\author[15]{Nicolas Tessore\orcidlink{0000-0002-9696-7931}}
\author[7,16,12]{Isaac Tutusaus\orcidlink{0000-0002-3199-0399}}
\author[2]{Casper Vedder\orcidlink{0009-0007-6341-4648}}
\author[ ]{, on behalf of the Euclid Consortium}

\affil[1]{Waterloo Centre for Astrophysics, University of Waterloo, Canada}
\affil[2]{Leiden Observatory, Leiden University, Netherlands}
\affil[3]{Centre for Astrophysics Research, University of Hertfordshire, United Kingdom}
\affil[4]{Scientific Information, German Aerospace Center (DLR), Germany}
\affil[5]{INAF - Osservatorio Astronomico di Trieste, Italy}
\affil[6]{INAF - Osservatorio Astronomico di Brera, Italy}
\affil[7]{Institute of Space Sciences (ICE, CSIC), Spain}
\affil[8]{INAF - IASF Milano, Italy}
\affil[9]{Astronomical Institute (AIRUB), Ruhr University Bochum, Germany}
\affil[10]{Department of Physics, University of Oxford, United Kingdom}
\affil[11]{Dipartimento di Fisica "Aldo Pontremoli", Università degli Studi di Milano, Italy}
\affil[12]{IRAP, Université de Toulouse, France}
\affil[13]{Osservatorio Astronomico di Roma, Italy}
\affil[14]{CCAPP, The Ohio State University, USA}
\affil[15]{Mullard Space Science Laboratory, University College London, United Kingdom}
\affil[16]{Institut d'Estudis Espacials de Catalunya (IEEC), Spain}


\maketitle

\marginpar{
  \begin{flushleft}
  \vspace{-10pt}
  \sffamily\small

  {\bfseries DOI:} \href{https://doi.org/DOI\_TBD}{\color{linky}{DOI TBD}}

  \vspace{2mm}

  {\bfseries Software}
  \begin{itemize}
    \setlength\itemsep{0em}
    \item \href{N/A}{\color{linky}{Review}} \ExternalLink
    \item \href{https://github.com/cloe-org/cloelike}{\color{linky}{Repository}} \ExternalLink
    \item \href{https://zenodo.org/record/TBD}{\color{linky}{Archive (Zenodo)}} \ExternalLink
  \end{itemize}

  \vspace{2mm}
  \par\noindent\hrulefill\par
  \vspace{2mm}

  {\bfseries Editor:} \href{https://example.com}{Pending editor} \ExternalLink \\
  \vspace{1mm}
  {\bfseries Reviewers:}
  \begin{itemize}
    \setlength\itemsep{0em}
    \item \href{https://github.com/pending}{@pending}
  \end{itemize}

  \vspace{2mm}
  {\bfseries Submitted:} N/A \\
  {\bfseries Published:} N/A

  \vspace{2mm}
  {\bfseries License}\\
  Authors retain copyright and release the work under CC BY 4.0
  (\href{http://creativecommons.org/licenses/by/4.0/}{\color{linky}{link}}).
  \end{flushleft}
}

\vspace{.2cm}

\vspace{.5cm}

\section{Summary}
\texttt{cloelike}, available at \href{https://github.com/cloe-org/cloelike}{cloe-org/cloelike}, is a Python package providing modular, composable Gaussian likelihood classes for the main cosmological large-scale structure observables targeted by the ESA \emph{Euclid} space mission. It is a core component of the \texttt{CLOE} (Cosmology Likelihood for Observables in Euclid) ecosystem and interfaces directly with \texttt{cloelib} for theoretical predictions and \texttt{euclidlib} for reading official \emph{Euclid} data products. The package implements Gaussian likelihoods covering harmonic angular power spectra and real-space two-point correlation functions for weak lensing (WL), photometric galaxy clustering (GCph), and Galaxy--Galaxy Lensing (GGL) in all joint probe combinations (3$\times$2pt, 2$\times$2pt), as well as spectroscopic full-shape power spectrum multipoles, and baryonic Acoustic oscillations (BAO). \texttt{cloelike} is actively used in internal Euclid Consortium analyses and is openly released to support community validation and reproducibility.

\hypertarget{statement-of-need}{%
\section{Statement of need}\label{statement-of-need}}

The ESA \emph{Euclid} mission is delivering high-precision photometric and spectroscopic observations aimed at probing the nature of dark energy and dark matter, as well as constraining the origin of primordial density perturbations in the early Universe. Achieving these scientific goals requires confronting (extended) cosmological models with the data through robust, validated, and reproducible likelihood frameworks for statistical inference~\citep{Euclid:2024, EP-CLOE3}. Despite the wide ecosystem of cosmological inference tools, such as \texttt{CosmoSIS}~\citep{2015-Cosmosis} and \texttt{Cobaya} \citep{2021-Cobaya}, most publicly available likelihood implementations remain probe-specific or lack the modularity and native data-format compatibility needed for fully integrated end-to-end \emph{Euclid} analyses.

\texttt{cloelike} addresses this limitation by providing a minimal and unified Python interface dedicated exclusively to computing a $\chi^2$ statistic. It operates on a data vector, covariance matrix supplied externally, and theoretical predictions supplied by \texttt{cloelib}, all consistently structured within the \texttt{euclidlib}\footnote{\href{https://github.com/euclidlib/euclidlib}{https://github.com/euclidlib/euclidlib}} data format. This strict separation of concerns, combined with seamless integration, ensures interoperability while avoiding duplication of functionality already provided by external inference frameworks or posterior samplers.

Concretely, \texttt{cloelike} evaluates the standard multivariate Gaussian likelihood through the quadratic form
\begin{equation}
-2\ln \mathcal{L}
=
(\mathbf{d}-\mathbf{t})^{\mathrm T}
\mathbf{C}^{-1}
(\mathbf{d}-\mathbf{t})
+
b,
\end{equation}
where $\mathbf{d}$ denotes the data vector, $\mathbf{t}$ the theoretical prediction, $\mathbf{C}$ the covariance matrix, and $b$ is an additive bias. This quantity fully specifies the Gaussian log-likelihood up to an additive normalization constant and constitutes the essential input required by most inference engines.

By design, \texttt{cloelike} is a lightweight and modular package that focuses exclusively on likelihood evaluation. It can be seamlessly interfaced with established Bayesian frameworks such as \texttt{Cobaya} or \texttt{CosmoSIS}, as well as integrated into custom sampling pipelines that operate on $\chi^2$ evaluations. Core tasks such as parameter exploration, sampling strategies, and model management are deliberately delegated to external tools, ensuring that \texttt{cloelike} remains dedicated to providing a consistent, validated, and reproducible likelihood computation.

This modular architecture enforces a clear separation between data handling, theoretical prediction, and statistical inference. In turn, \texttt{cloelike} lowers the barrier for the community to reproduce official \emph{Euclid} results, while enabling flexible exploration of new systematics, extended cosmological models, and probe combinations. The package is actively used within the Euclid Consortium, where it serves as the reference likelihood implementation and represents a natural evolution of the former \texttt{CLOE} software~\citep{EP-CLOE2, EP-CLOE1, EP-CLOE4}.

\section{State of the Field}
\label{sec:state_of_field}
The ecosystem of cosmological likelihood and inference frameworks is both extensive and heterogeneous. Widely used examples include \texttt{CosmoSIS}, which has been broadly adopted across the community and in major galaxy surveys such as the \emph{Dark Energy Survey} (DES); \texttt{Cobaya}, favored among theoretical cosmologists for its flexibility in combining multiple data sets and interfacing with different Boltzmann solvers; \texttt{CosmoLike}~\citep{CosmoLike}; and \texttt{Firecrown}\footnote{\href{https://github.com/LSSTDESC/firecrown}{https://github.com/LSSTDESC/firecrown}}, among others. Furthermore, survey-specific pipelines and wrappers have been developed, such as \texttt{CosmoPipe}\footnote{\href{https://github.com/AngusWright/CosmoPipe}{https://github.com/AngusWright/CosmoPipe}} for \emph{Kilo Degree Survey} (KiDS) analyses. These frameworks typically provide end-to-end Bayesian analysis environments, combining theoretical predictions, data handling, likelihood evaluation, and interfaces to sampling engines within a unified software structure.

In contrast, \texttt{cloelike} adopts a deliberately modular and simpler philosophy, closer in spirit to self-contained likelihood implementations such as the \emph{Planck 2018} \texttt{plik} likelihood~\citep{Planck2018}. Rather than embedding the full inference pipeline, \texttt{cloelike} focuses exclusively on providing a validated and consistent likelihood layer with native support for \emph{Euclid} data products across all primary probes. In this way, it complements existing frameworks: it can be seamlessly integrated as a drop-in likelihood module within \texttt{CosmoSIS} or \texttt{Cobaya} workflows, allowing users to exploit advanced sampling capabilities while relying on a robust, probe-complete likelihood implementation. At the same time, its minimal interface based solely on $\chi^2$ evaluation, makes it readily compatible with any external sampler or inference tool that operates on likelihood values, enabling flexible integration within a wide range of analysis pipelines [e.g.: \texttt{Nautilus} \citep{nautilus}, \texttt{dynesty} \citep{Higson2019}, \texttt{emcee} \citep{emcee}, and others].

\section{Software Design}
\label{sec:software}
\texttt{cloelike} supports likelihoods already prepared for photometric harmonic and configuration-space summary statistics (the latter including two-point correlation functions and COSEBIs), as well as spectroscopic galaxy clustering observables, both in their full-shape and BAO formulations, with the possibility of constructing a joint likelihood combining the two spectroscopic probes.

The internal architecture of \texttt{cloelike} follows a modular design based on a strict separation of concerns. For photometric probes, which share a common statistical structure, all operations related to data-vector ingestion, masking, scale cuts, covariance loading, precision-matrix handling, and consistency checks are implemented in a shared base class\footnote{For spectroscopic galaxy clustering, given its self-standing nature but different from photometric probes, a common class is not neccesary.}. This common layer provides a unified interface for handling observational inputs independently of the specific summary statistic being analysed. On top of this shared infrastructure, the computation of the theoretical observables is delegated to independent probe-specific mixins. Each mixin encapsulates only the logic required to compute a given observable—for example, harmonic-space angular power spectra, configuration-space correlation functions, or COSEBIs—while inheriting the shared data and covariance machinery from the base class. This approach avoids code duplication, ensures consistency across probes, and makes the addition of new observables straightforward, since only the corresponding theoretical prediction module needs to be implemented.

This architecture closely mirrors the protocol-oriented modularity adopted in \texttt{cloelib}, with which \texttt{cloelike} interfaces directly. In practice, \texttt{cloelike} acts as the likelihood layer responsible for statistical evaluation, while \texttt{cloelib} provides the theoretical predictions through interchangeable protocols. The result is a flexible and extensible framework in which likelihood definitions and theory computations remain cleanly decoupled, facilitating maintenance, validation, and future extensions to new cosmological probes.

Building on this modular foundation, \texttt{cloelike} enables a high degree of flexibility in composing likelihoods across probes and summary statistics. Each likelihood instance is effectively constructed by combining a data container, a covariance representation, and one or more observable providers, allowing users to tailor configurations without modifying the underlying infrastructure. This composability is particularly advantageous in the context of multi-probe survey analyses, where consistency across probes must be maintained while retaining the ability to evolve individual components independently.

A key feature of the framework is its seamless integration with theory predictions provided by \texttt{cloelib} using Python protocols. Through a well-defined interface, \texttt{cloelike} retrieves theoretical predictions for all supported observables, ensuring consistency in cosmological parameter definitions, nuisance modeling, and numerical settings. This tight coupling minimizes duplication of functionality and code, and reduces the risk of inconsistencies between likelihood and theory layers. Moreover, it allows \texttt{cloelike} to benefit directly from any future developments in \texttt{cloelib}. This includes, for example, ongoing improvements in non-linear modeling, such as galaxy bias, redshift-space distortions, and intrinsic alignment, as well as extensions to beyond-$\varLambda$CDM scenarios.

From an implementation perspective, Python mixins in the photometric case provide an elegant yet powerful mechanism to extend the framework for likelihood combinations, such as 3$\times$2pt and 2$\times$2pt. New probes or summary statistics can be incorporated by defining additional mixins that implement the required observable computations and data-vector mappings, without altering the core likelihood machinery. This promotes rapid prototyping and facilitates the inclusion of novel observables, such as cross-correlations with the cosmic microwave background, which is essential for future data releases and methodological developments within the Euclid Consortium.

All \texttt{cloelike} likelihoods, independently of the probe or data space considered, are initialised from two Python dictionaries that provide the required data products and analysis specifications. The \texttt{data} dictionary contains the data vector, auxiliary data products, and the corresponding covariance matrix, while the \texttt{settings} dictionary defines analysis choices such as scale cuts for each redshift bin in tomographic analyses.

In addition, each likelihood requires the appropriate \texttt{cloelib} protocol-compatible theory classes. Photometric likelihoods require classes for \texttt{Background} and \texttt{Perturbations}; spectroscopic galaxy-clustering full-shape likelihoods require \texttt{Background} and \texttt{SpectroPower}; and BAO likelihoods require only a \texttt{Background} class.

Moreover, \texttt{cloelike} is designed with performance and scalability in mind. Vectorized operations, caching of intermediate quantities (inverse of the covariance matrix), and compatibility with parallel sampling codes ensure that the likelihood evaluations remain efficient even for high-dimensional parameter spaces. The framework is also compatible with standard sampling and optimization tools, enabling its deployment in both Bayesian inference and frequentist pipelines.

Finally, the architecture naturally supports joint analyses across multiple probes. By construction, shared parameters spaces, as long as there is no cross-covariances between the probes, can be consistently handled, allowing \texttt{cloelike} to perform coherent multi-probe likelihood evaluations by means of summing log-likelihood evaluations. This capability is central to extracting the full scientific potential of forthcoming survey data, where combined constraints from weak lensing, galaxy clustering, and other observables will play a decisive role in testing cosmological models.



\section{Usage Examples}\label{usage-examples}
In this section, we illustrate the initialization and evaluation of a single $\chi^2$ for two representative likelihoods implemented in \texttt{cloelike}. These examples expect to ingest data vectors, covariance matrices, and other data products (i.e: galaxy redshift distributions) in the \texttt{euclidlib} format, as well to specify scale cuts on those data vectors. 

\subsection{Harmonic photometric likelihoods}\label{photo-likelihoods}
In this example, we evaluate the log-likelihood for cosmic shear angular power spectra. For photometric likelihood analyses, \texttt{cloelike} is designed to ingest covariance matrices ordered by "probe", "redshift bin", and "multipole". This convention corresponds to the native output format of the \textit{Euclid} covariance matrix software \texttt{Spaceborne}\footnote{\href{https://github.com/davidesciotti/Spaceborne}{https://github.com/davidesciotti/Spaceborne}}.

\begin{lstlisting}[language=python]
# ------------------------------------------------------------------
# Step 1: Construct the data dictionary expected by cloelike
# ------------------------------------------------------------------
# This function assembles the inputs required by the likelihood:
# - Angular power spectra (cls_data)
# - Multipole range (ells)
# - Galaxy Redshift distribution (myz, my_dndz_she_norm)
# - Covariance matrix (cov)
# - Mixing matrix (mixmats), accounting for mode coupling

def build_data_WL(ell_key, cov):
    return {
        'cells': cls_data,
        'ells': cls_data[ell_key].ell,
        'z_arr': myz,
        'cov': cov,
        'mixmat': mixmats,
        'dndz_she': my_dndz_she_norm, # must be normalised
    }


# ------------------------------------------------------------------
# Step 2: Define analysis settings dictionary expected by cloelike
# ------------------------------------------------------------------
# These control how the data vector is compressed and which scales
# are included in the likelihood evaluation.

def build_settings():
    scale_cuts = {key: [5, 3000] for key in cls_data}
    return {
        'n_ell_bins': 32,        # Number of multipole bins
        'scale_cuts': scale_cuts # Minimum and maximum ell per probe
    }


# ------------------------------------------------------------------
# Step 3: Instantiate the dataset
# ------------------------------------------------------------------
# We select the shear-shear (SHE, SHE) auto-correlation as the
# observable defining the weak lensing probe.

data_WL = build_data_WL(('SHE', 'SHE', 1, 1), covmat_WL)
settings_WL = build_settings()


# ------------------------------------------------------------------
# Step 4: Initialise the likelihood
# ------------------------------------------------------------------
# The likelihood combines:
# - Background cosmology from cloelib (CAMBBackground)
# - Linear perturbations from cloelib (HMemuLinearPerturbations)
# - Non-linear corrections from cloelib (HMemuNonLinearPerturbations)
#
# Initialisation includes precomputations and typically dominates
# the one-time setup cost as it also prepares the data

like_WL = EuclidLikelihood_WL(
    data=data_WL,
    settings=settings_WL,
    Background=CAMBBackground,
    LinPerturbations=HMemuLinearPerturbations,
    NonLinPerturbations=HMemuNonLinearPerturbations,
    mode='coupled'
)


# ------------------------------------------------------------------
# Step 5: Define fiducial cosmological and nuisance parameters
# ------------------------------------------------------------------
# Evaluating the likelihood at this point should provide a non-zero value

default_pars = {
    # Cosmological parameters
    'H0': 70,
    'Omega_cdm0': 0.25,
    'Omega_b0': 0.05,
    'ns': 0.96,
    'As': 2e-9,
    'w0': -1,
    'wa': 0,
    'Omega_k0': 0,
    'mnu': 0.06,
    'gamma_MG': 0.545,
    'N_mnu': 1,
    'log10TAGN': 7.8,

    # Intrinsic alignment model
    'AIA': 1.72,
    'CIA': 0.0134,
    'EtaIA': -0.41,

    # Shear calibration systematics
    'multiplicative_bias_1': 0.0,
    'multiplicative_bias_2': 0.0,
    'multiplicative_bias_3': 0.0,
    'multiplicative_bias_4': 0.0,
    'multiplicative_bias_5': 0.0,
    'multiplicative_bias_6': 0.0,

    # Photometric redshift uncertainties (shear)
    'dz_shear_1': 0.0,
    'dz_shear_2': 0.0,
    'dz_shear_3': 0.0,
    'dz_shear_4': 0.0,
    'dz_shear_5': 0.0,
    'dz_shear_6': 0.0,

    # Photometric redshift uncertainties (shear)
    'width_shear_1': 0.0,
    'width_shear_2': 0.0,
    'width_shear_3': 0.0,
    'width_shear_4': 0.0,
    'width_shear_5': 0.0,
    'width_shear_6': 0.0,
}


# ------------------------------------------------------------------
# Step 6: Evaluate the log-likelihood
# ------------------------------------------------------------------
# This step computes the agreement between the theoretical prediction
# (given the parameters above) and the observed data vector.


loglike = like_WL.loglike(default_pars)

\end{lstlisting}

\subsection{Spectroscopic galaxy clustering full-shape}\label{spectro-likelihoods}
In this example, we compute a log-likelihood evaluation for spectroscopic galaxy clustering full-shape.

\begin{lstlisting}[language=python]
# ------------------------------------------------------------------
# Step 1: Define the spectroscopic redshift bins
# ------------------------------------------------------------------
# These labels identify the redshift bins used when reading the
# corresponding Euclid data-vector, covariance, and mixing-matrix files.

redshifts = [1.0, 1.2, 1.4, 1.65]
labels = [str(z).strip('0') for z in redshifts]


# ------------------------------------------------------------------
# Step 2: Read the spectroscopic GC data products
# ------------------------------------------------------------------
# euclidlib allows multiple files to be read at once by passing the
# redshift-bin labels to the reader functions.

datavec = power_spectrum_multipoles(
    'mps_pk_GCspectro_comet_EFT_z{}.fits',
    *labels
)

covariance = power_spectrum_multipole_covariance(
    'cov_pk_Gauss_GCspectro_comet_EFT_z{}_2500deg2.fits',
    *labels
)

mixing = power_spectrum_multipole_mixing_matrix(
    'mixmat_pk_GCspectro_identity_z{}.fits',
    *labels
)


# ------------------------------------------------------------------
# Step 3: Construct the data dictionary expected by cloelike
# ------------------------------------------------------------------
# The spectroscopic likelihood requires:
# - Power spectrum multipoles P_0(k), P_2(k), and P_4(k)
# - Effective k values
# - Number densities
# - Covariance matrices
# - Mixing matrices
# - Fiducial cosmology, required for AP distortions

data = {
    'GCspectro': {},
    'fiducial_cosmology': {}
}


# ------------------------------------------------------------------
# Step 4: Store the fiducial cosmology
# ------------------------------------------------------------------
# We extract the fiducial cosmology from the data vector and complete
# the parameter dictionary with the quantities required by cloelib.

dv = datavec[("SPE", "SPE", 0, 0)]

data['fiducial_cosmology']['H0'] = dv.fiducial_cosmology['H0']
data['fiducial_cosmology']['Omega_cdm0'] = (
    dv.fiducial_cosmology['Omega_m0']
    - dv.fiducial_cosmology['Omega_b0']
)
data['fiducial_cosmology']['Omega_b0'] = dv.fiducial_cosmology['Omega_b0']
data['fiducial_cosmology']['Omega_k0'] = dv.fiducial_cosmology['Omega_k0']
data['fiducial_cosmology']['mnu'] = 0.0
data['fiducial_cosmology']['N_mnu'] = 0
data['fiducial_cosmology']['w0'] = dv.fiducial_cosmology['w0']
data['fiducial_cosmology']['wa'] = 0.0
data['fiducial_cosmology']['ns'] = dv.fiducial_cosmology['ns']
data['fiducial_cosmology']['As'] = 2.1e-9
data['fiducial_cosmology']['gamma_MG'] = 0.545


# ------------------------------------------------------------------
# Step 5: Convert units consistently
# ------------------------------------------------------------------
# The data vectors are stored in Mpc/h units, while cloelib currently
# works internally in Mpc units. We therefore rescale k, P(k), the
# covariance, and number densities using the fiducial value of h.

fid_h = data['fiducial_cosmology']['H0'] / 100.0

k_fac = fid_h
pk_fac = 1.0 / fid_h**3
cov_fac = 1.0 / fid_h**6


# ------------------------------------------------------------------
# Step 6: Fill the redshift-dependent GCspectro data blocks
# ------------------------------------------------------------------
# For each redshift bin, we store the multipoles, covariance matrix,
# number density, and mixing matrix in the structure expected by
# EuclidLikelihood_GCspectro_Pls.

for ii, z in enumerate(labels):

    data['GCspectro'][z] = {}

    dv_inst = datavec[('SPE', 'SPE', ii, ii)]
    cv_inst = covariance[('SPE', 'SPE', ii, ii)]
    mm_inst = mixing[('SPE', 'SPE', ii, ii)]

    # Number density
    data['GCspectro'][z]['nbar'] = dv_inst.nbar * fid_h**3

    # Power spectrum multipoles
    data['GCspectro'][z]['k'] = dv_inst.keff * k_fac
    data['GCspectro'][z]['pk0'] = dv_inst.multipoles[0] * pk_fac
    data['GCspectro'][z]['pk2'] = dv_inst.multipoles[2] * pk_fac
    data['GCspectro'][z]['pk4'] = dv_inst.multipoles[4] * pk_fac

    # Covariance matrix
    full_matrix = np.block([
        [cv_inst.covariance[f'ELL_{i}-{j}'] for j in [0, 2, 4]]
        for i in [0, 2, 4]
    ])
    data['GCspectro'][z]['cov'] = full_matrix * cov_fac

    # Mixing matrix
    mixing_matrix_dict = {
        'kout': mm_inst.kout * k_fac
    }

    mixing_matrix_dict.update({
        f'kin{ell}': mm_inst.kin[ell] * k_fac
        for ell in [0, 2, 4]
    })

    mixing_matrix_dict.update({
        f'W{i}{j}': mm_inst.mixing[f'ELL_{i}-{j}'].squeeze()
        for i, j in product([0, 2, 4], repeat=2)
    })

    data['GCspectro'][z]['mixing_matrix'] = mixing_matrix_dict


# ------------------------------------------------------------------
# Step 7: Define the analysis settings
# ------------------------------------------------------------------
# Scale cuts are initially specified in Mpc/h units for each multipole
# and redshift bin. They are converted below to Mpc units to remain
# consistent with cloelib.

settings = {
    'scale_cuts': {
        'GCspectro': {
            'bin1': {
                'ell0': [0.0, 0.20],
                'ell2': [0.0, 0.15],
                'ell4': [0.0, 0.15],
            },
            'bin2': {
                'ell0': [0.0, 0.25],
                'ell2': [0.0, 0.20],
                'ell4': [0.0, 0.20],
            },
            'bin3': {
                'ell0': [0.0, 0.25],
                'ell2': [0.0, 0.20],
                'ell4': [0.0, 0.20],
            },
            'bin4': {
                'ell0': [0.0, 0.30],
                'ell2': [0.0, 0.25],
                'ell4': [0.0, 0.25],
            },
        }
    }
}

settings = {
    'scale_cuts': {
        'GCspectro': {
            bin_key: {
                ell_key: [v * fid_h for v in values]
                for ell_key, values in bin_values.items()
            }
            for bin_key, bin_values
            in settings['scale_cuts']['GCspectro'].items()
        }
    }
}


# ------------------------------------------------------------------
# Step 8: Initialise the likelihood
# ------------------------------------------------------------------
# The likelihood combines:
# - Background cosmology from cloelib
# - Spectroscopic galaxy clustering predictions from CometEFT
#
# The likelihood receives the prepared Euclid data products and computes
# the spectroscopic GC power-spectrum likelihood.

like_spec = EuclidLikelihood_GCspectro_Pls(
    data=data,
    settings=settings,
    Background=CAMBBackground,
    SpectroPower=CometEFT_SpectroPower,
)


# ------------------------------------------------------------------
# Step 9: Define cosmological and nuisance parameters
# ------------------------------------------------------------------
# The nuisance parameters are provided as arrays, with one value per
# spectroscopic redshift bin.

parameters = {
    # Cosmological parameters
    'H0': 67.0,
    'Omega_cdm0': 0.27,
    'Omega_b0': 0.049,
    'Omega_k0': 0.0,
    'mnu': 0.0,
    'N_mnu': 0,
    'w0': -1.0,
    'wa': 0.0,
    'ns': 0.96,
    'As': 2.1e-9,
    'gamma_MG': 0.545,

    # Galaxy bias and EFT nuisance parameters
    'b1': np.array([1.412, 1.769, 2.039, 2.496]),
    'b2': np.array([0.695, 0.870, 1.162, 2.010]),
    'bG2': np.array([-0.156, -0.299, -0.400, -0.555]),
    'bGam3': np.array([0.323, 0.621, 0.827, 1.137]),
    'c0': np.array([30.948, 37.116, 36.738, 53.627]),
    'c2': np.array([46.233, 53.071, 48.626, 60.962]),
    'c4': np.array([10.057, 10.385, 8.643, 8.711]),
    'cnlo': np.array([0.0, 0.0, 0.0, 0.0]),

    # Shot-noise and observational nuisance parameters
    'NP0': np.array([1.056, 1.152, 1.144, 1.309]),
    'NP20': np.array([0.0, 0.0, 0.0, 0.0]),
    'NP22': np.array([0.0, 0.0, 0.0, 0.0]),
    'fout': np.array([0.0, 0.0, 0.0, 0.0]),
    'sigmaz': np.array([0.0, 0.0, 0.0, 0.0]),
}


# ------------------------------------------------------------------
# Step 10: Evaluate the log-likelihood
# ------------------------------------------------------------------
# This computes the agreement between the spectroscopic GC data vector
# and the theoretical prediction for the parameter values above.

loglike = like_spec.loglike(parameters)
\end{lstlisting}


\hypertarget{documentation}{%
\section{Documentation}\label{documentation}}

Comprehensive documentation for \texttt{cloelike} is available at
\href{https://cloe-org.github.io/cloelike/dev/home/}{cloe-org.github.io/cloelike/dev/home/}.
The documentation includes detailed API references, installation
instructions, explanations about the software structure, and guides for
integrating \texttt{cloelike} into your analysis workflows.

For practical examples, example scripts, and interactive tutorials,
visit the
\href{https://github.com/cloe-org/playground}{cloe-org/playground}
repository, which hosts a collection of Jupyter notebooks showcasing
typical use cases and advanced features. In particular, \href{https://github.com/cloe-org/playground/tree/main/tutorials/likelihood}{cloe-org/playground/tutorials/likelihood} shows examples on how to run single $\chi^2$ evaluations. 

\texttt{cloelike} is prepared to interface with photometric and spectroscopic
observables data formats compliant with
\texttt{euclidlib}\footnote{\href{https://euclidlib.readthedocs.io/en/latest/}{https://euclidlib.readthedocs.io/en/latest/}}
formats, using
\texttt{cosmolib}\footnote{\href{https://github.com/astro-ph/cosmolib}{https://github.com/astro-ph/cosmolib}}
dataclasses.

\section{Availability}
\noindent
\textbf{Source:} \href{https://github.com/cloe-org/cloelib}{github.com/cloe-org/cloelike} \\
\textbf{License:} MIT. \\
\textbf{Install (PyPI):} \texttt{pip\,\, install\,\, cloelike} \\
\textbf{Documentation:} \href{ https://your_package.readthedocs.com}{https://cloe-org.github.io/cloelike/dev/home/}\\
\textbf{Examples:}
\href{https://github.com/cloe-org/playground}{github.com/cloe-org/playground}\\


\section*{Acknowledgments}

We thank the broader CLOE software development team for foundational work that motivated this
library. The scientific development of cloe-org is coordinated through the Joint Cosmology Key Project one (DR1-KP-JC-1) of the Euclid Consortium, led by C. Carbone, M. Crocce, and I. Tutusaus. In this context, \texttt{cloe-org} has been adopted as the default analysis code of the Euclid Consortium. G.C.H. acknowledges that this project is part of the project
UNICORN with file number VI.Veni.242.110 of the research programme
Talent Programme Veni Science domain 2024 which is (partly) financed by
the Dutch Research Council (NWO) under the grant
https://doi.org/10.61686/ZCPQI32997. M.B. acknowledges support from the
Natural Sciences and Engineering Research Council of Canada (NSERC).
C.M. is supported by the Agenzia Spaziale Italiana project ``Attività
scientifica per la missione Euclid -- fase E ACCORDO ATTUATIVO
n.~2024-10-HH.0.''. I.T. acknowledges support form the Spanish Ministerio de Ciencia, Innovaci\'on y Universidades, projects PID2022-141079NB, PID2022-138896NB; the European Research Executive Agency HORIZON-MSCA-2021-SE-01 Research and Innovation programme under the Marie Sk\l odowska-Curie grant agreement number 101086388 (LACEGAL) and the programme Unidad de Excelencia Mar\'{\i}a de Maeztu, project CEX2020-001058-M. 

We acknowledge EuroHPC Joint Undertaking for awarding the project ID
EHPC-EXT-2024E02-083 access to Leonardo hosted by CINECA, Italy. We
acknowledge the use of Spanish Supercomputing Network (RES) resources
provided by the Barcelona Supercomputing Center (BSC) in MareNostrum 5
under allocations AECT-2024-3-0020, 2025-1-0045, 2025-2-0046,
2025-3-0036. We acknowledge support from the European Research Council
(ERC) under the European Union's Horizon 2020 research and innovation
program with Grant agreement No.~101053992 for computational resources.

\AckEC 

\section*{Author Contributions}

In accordance with JOSS guidelines, we describe individual contributions below. Authors are listed in alphabetical order. All Tier 1 authors (Bonici, Cañas-Herrera, Carrilho, Casas, Moretti, Pezzotta) are core maintainers of the \textbf{cloe-org} organisation, responsible for the long-term sustainability of \texttt{cloelike}, the review of pull requests, and leadership of technical discussions.
\begin{itemize}
\tightlist
\item
  \textbf{M. Bonici}: Review of software and design.
\item
  \textbf{G. Cañas-Herrera}: Repository set-up and overall software architecture; implementation of \texttt{PhotoLikelihoodBase} including data ingestion, scale-cut masking, covariance inversion, geometric rebinning of $C_\ell$ data vectors, and \texttt{lru\_cache} optimisation; mixin composition framework; CI pipeline configuration; pre-commit and code-quality tooling; issue and pull-request templates; README, documentation infrastructure, and community contribution tracking (\texttt{all-contributors}); \texttt{pyproject.toml} versioning and release workflows; BAO unit tests. Resources, writing - original draft, visualization, project administration.
\item
  \textbf{P. Carrilho}: Implementation of systematic nuisance parameters for photometric tracers (multiplicative shear bias, photometric redshift shifts and width parameters); coupled/uncoupled mode flag for pseudo-$C_\ell$ likelihood with mixing matrices; photometric likelihood refinements and notebook tutorials.
\item
  \textbf{S. Casas}: Review of design and ideas.
\item
  \textbf{C. Moretti}:  Implementation of \texttt{EuclidLikelihood\_BAO} (BAO $\alpha$-parameter likelihood); spectroscopic likelihood bug fixes; unit tests for BAO likelihood; repository housekeeping.
\item
  \textbf{A. Pezzotta}: Core implementation of \texttt{EuclidLikelihood\_GCspectro\_Pls} (full-shape spectroscopic power spectrum multipoles); analytical marginalisation over linear bias parameters; spectroscopic data-vector and covariance-matrix handling; unit tests for spectroscopic likelihood; homogenisation of photometric and spectroscopic likelihood APIs.
\end{itemize}

C. Carbone, M. Crocce, and I. Tutusaus: Project administration, Supervision, Validation. The contributions of all remaining authors have been tracked using the
\href{https://github.com/all-contributors/all-contributors}{all-contributors}
bot, following the specification of the same name. A full, categorised
breakdown of each contributor's role---including code, documentation,
testing, ideas, project management, and more---is available in the
\texttt{README} of the \texttt{cloelike} repository, fully detailed
within the \texttt{cloelike} docs.

\bibliographystyle{mnras}
\bibliography{main_joss}

@article{cloelib,
    author = {cloe-org maintainers},
    title = {\texttt{cloelib}: A Flexible Python Library for Computing Cosmological
Observables in the Euclid Era},
    journal = JOSS,
    year = 2026
}

@article{Higson2019,
  author    = {Higson, Edward and Handley, Will and Hobson, Michael P. and Lasenby, Anthony},
  title     = {Dynamic nested sampling: an improved algorithm for parameter estimation and evidence calculation},
  journal   = {Statistics and Computing},
  year      = {2019},
  volume    = {29},
  pages     = {891--913},
  doi       = {10.1007/s11222-018-9844-0}
}

@ARTICLE{emcee,
       author = {{Foreman-Mackey}, Daniel and {Hogg}, David W. and {Lang}, Dustin and {Goodman}, Jonathan},
        title = "{emcee: The MCMC Hammer}",
      journal = {\pasp},
         year = 2013,
        month = mar,
       volume = {125},
       number = {925},
        pages = {306},
          doi = {10.1086/670067},
archivePrefix = {arXiv},
       eprint = {1202.3665},
 primaryClass = {astro-ph.IM},
       adsurl = {https://ui.adsabs.harvard.edu/abs/2013PASP..125..306F}
}

@ARTICLE{Planck2018,
       author = {{Planck Collaboration} and {Aghanim}, N. and {Akrami}, Y. and {Ashdown}, M. and {Aumont}, J. and {Baccigalupi}, C. and {Ballardini}, M. and {Banday}, A.~J. and {Barreiro}, R.~B. and {Bartolo}, N. and {Basak}, S. and {Benabed}, K. and {Bernard}, J.-P. and {Bersanelli}, M. and {Bielewicz}, P. and {Bock}, J.~J. and {Bond}, J.~R. and {Borrill}, J. and {Bouchet}, F.~R. and {Boulanger}, F. and {Bucher}, M. and {Burigana}, C. and {Butler}, R.~C. and {Calabrese}, E. and {Cardoso}, J.-F. and {Carron}, J. and {Casaponsa}, B. and {Challinor}, A. and {Chiang}, H.~C. and {Colombo}, L.~P.~L. and {Combet}, C. and {Crill}, B.~P. and {Cuttaia}, F. and {de Bernardis}, P. and {de Rosa}, A. and {de Zotti}, G. and {Delabrouille}, J. and {Delouis}, J.-M. and {Di Valentino}, E. and {Diego}, J.~M. and {Dor{\'e}}, O. and {Douspis}, M. and {Ducout}, A. and {Dupac}, X. and {Dusini}, S. and {Efstathiou}, G. and {Elsner}, F. and {En{\ss}lin}, T.~A. and {Eriksen}, H.~K. and {Fantaye}, Y. and {Fernandez-Cobos}, R. and {Finelli}, F. and {Frailis}, M. and {Fraisse}, A.~A. and {Franceschi}, E. and {Frolov}, A. and {Galeotta}, S. and {Galli}, S. and {Ganga}, K. and {G{\'e}nova-Santos}, R.~T. and {Gerbino}, M. and {Ghosh}, T. and {Giraud-H{\'e}raud}, Y. and {Gonz{\'a}lez-Nuevo}, J. and {G{\'o}rski}, K.~M. and {Gratton}, S. and {Gruppuso}, A. and {Gudmundsson}, J.~E. and {Hamann}, J. and {Handley}, W. and {Hansen}, F.~K. and {Herranz}, D. and {Hivon}, E. and {Huang}, Z. and {Jaffe}, A.~H. and {Jones}, W.~C. and {Keih{\"a}nen}, E. and {Keskitalo}, R. and {Kiiveri}, K. and {Kim}, J. and {Kisner}, T.~S. and {Krachmalnicoff}, N. and {Kunz}, M. and {Kurki-Suonio}, H. and {Lagache}, G. and {Lamarre}, J.-M. and {Lasenby}, A. and {Lattanzi}, M. and {Lawrence}, C.~R. and {Le Jeune}, M. and {Levrier}, F. and {Lewis}, A. and {Liguori}, M. and {Lilje}, P.~B. and {Lilley}, M. and {Lindholm}, V. and {L{\'o}pez-Caniego}, M. and {Lubin}, P.~M. and {Ma}, Y.-Z. and {Mac{\'\i}as-P{\'e}rez}, J.~F. and {Maggio}, G. and {Maino}, D. and {Mandolesi}, N. and {Mangilli}, A. and {Marcos-Caballero}, A. and {Maris}, M. and {Martin}, P.~G. and {Mart{\'\i}nez-Gonz{\'a}lez}, E. and {Matarrese}, S. and {Mauri}, N. and {McEwen}, J.~D. and {Meinhold}, P.~R. and {Melchiorri}, A. and {Mennella}, A. and {Migliaccio}, M. and {Millea}, M. and {Miville-Desch{\^e}nes}, M.-A. and {Molinari}, D. and {Moneti}, A. and {Montier}, L. and {Morgante}, G. and {Moss}, A. and {Natoli}, P. and {N{\o}rgaard-Nielsen}, H.~U. and {Pagano}, L. and {Paoletti}, D. and {Partridge}, B. and {Patanchon}, G. and {Peiris}, H.~V. and {Perrotta}, F. and {Pettorino}, V. and {Piacentini}, F. and {Polenta}, G. and {Puget}, J.-L. and {Rachen}, J.~P. and {Reinecke}, M. and {Remazeilles}, M. and {Renzi}, A. and {Rocha}, G. and {Rosset}, C. and {Roudier}, G. and {Rubi{\~n}o-Mart{\'\i}n}, J.~A. and {Ruiz-Granados}, B. and {Salvati}, L. and {Sandri}, M. and {Savelainen}, M. and {Scott}, D. and {Shellard}, E.~P.~S. and {Sirignano}, C. and {Sirri}, G. and {Spencer}, L.~D. and {Sunyaev}, R. and {Suur-Uski}, A.-S. and {Tauber}, J.~A. and {Tavagnacco}, D. and {Tenti}, M. and {Toffolatti}, L. and {Tomasi}, M. and {Trombetti}, T. and {Valiviita}, J. and {Van Tent}, B. and {Vielva}, P. and {Villa}, F. and {Vittorio}, N. and {Wandelt}, B.~D. and {Wehus}, I.~K. and {Zacchei}, A. and {Zonca}, A.},
        title = "{Planck 2018 results. V. CMB power spectra and likelihoods}",
      journal = {\aap},
         year = 2020,
        month = sep,
       volume = {641},
          eid = {A5},
        pages = {A5},
          doi = {10.1051/0004-6361/201936386},
archivePrefix = {arXiv},
       eprint = {1907.12875},
 primaryClass = {astro-ph.CO},
       adsurl = {https://ui.adsabs.harvard.edu/abs/2020A&A...641A...5P}
}

@ARTICLE{2015-Cosmosis,
       author = {{Zuntz}, J. and {Paterno}, M. and {Jennings}, E. and {Rudd}, D. and {Manzotti}, A. and {Dodelson}, S. and {Bridle}, S. and {Sehrish}, S. and {Kowalkowski}, J.},
        title = "{CosmoSIS: Modular cosmological parameter estimation}",
      journal = {Astronomy and Computing},
         year = 2015,
        month = sep,
       volume = {12},
        pages = {45-59},
          doi = {10.1016/j.ascom.2015.05.005},
archivePrefix = {arXiv},
       eprint = {1409.3409},
 primaryClass = {astro-ph.CO},
       adsurl = {https://ui.adsabs.harvard.edu/abs/2015A&C....12...45Z}
}

@ARTICLE{2021-Cobaya,
       author = {{Torrado}, Jes{\'u}s and {Lewis}, Antony},
        title = "{Cobaya: code for Bayesian analysis of hierarchical physical models}",
      journal = {\jcap},
         year = 2021,
        month = may,
       volume = {2021},
       number = {5},
          eid = {057},
        pages = {057},
          doi = {10.1088/1475-7516/2021/05/057},
archivePrefix = {arXiv},
       eprint = {2005.05290},
 primaryClass = {astro-ph.IM},
       adsurl = {https://ui.adsabs.harvard.edu/abs/2021JCAP...05..057T}
}

@article{nautilus,
    author = {Lange, Johannes U},
    title = "{nautilus: boosting Bayesian importance nested sampling with deep learning}",
    journal = {Monthly Notices of the Royal Astronomical Society},
    volume = {525},
    number = {2},
    pages = {3181-3194},
    year = {2023},
    month = {08},
    doi = {10.1093/mnras/stad2441},
    url = {https://doi.org/10.1093/mnras/stad2441},
    eprint = {https://academic.oup.com/mnras/article-pdf/525/2/3181/51331635/stad2441.pdf},
}

@ARTICLE{Euclid:2024,
       author = {{Euclid Collaboration: Mellier}, Y.  and {Abdurro'uf} and {Acevedo Barroso}, J.~A. and {Ach{\'u}carro}, A. and {Adamek}, J. and {Adam}, R. and {Addison}, G.~E. and {Aghanim}, N. and {Aguena}, M. and {Ajani}, V. and {Akrami}, Y. and {Al-Bahlawan}, A. and {Alavi}, A. and {Albuquerque}, I.~S. and {Alestas}, G. and {Alguero}, G. and {Allaoui}, A. and {Allen}, S.~W. and {Allevato}, V. and {Alonso-Tetilla}, A.~V. and {Altieri}, B. and {Alvarez-Candal}, A. and {Alvi}, S. and {Amara}, A. and {Amendola}, L. and {Amiaux}, J. and {Andika}, I.~T. and {Andreon}, S. and {Andrews}, A. and {Angora}, G. and {Angulo}, R.~E. and {Annibali}, F. and {Anselmi}, A. and {Anselmi}, S. and {Arcari}, S. and {Archidiacono}, M. and {Aric{\`o}}, G. and {Arnaud}, M. and {Arnouts}, S. and {Asgari}, M. and {Asorey}, J. and {Atayde}, L. and {Atek}, H. and {Atrio-Barandela}, F. and {Aubert}, M. and {Aubourg}, E. and {Auphan}, T. and {Auricchio}, N. and {Aussel}, B. and {Aussel}, H. and {Avelino}, P.~P. and {Avgoustidis}, A. and {Avila}, S. and {Awan}, S. and {Azzollini}, R. and {Baccigalupi}, C. and {Bachelet}, E. and {Bacon}, D. and {Baes}, M. and {Bagley}, M.~B. and {Bahr-Kalus}, B. and {Balaguera-Antolinez}, A. and {Balbinot}, E. and {Balcells}, M. and {Baldi}, M. and {Baldry}, I. and {Balestra}, A. and {Ballardini}, M. and {Ballester}, O. and {Balogh}, M. and {Ba{\~n}ados}, E. and {Barbier}, R. and {Bardelli}, S. and {Baron}, M. and {Barreiro}, T. and {Barrena}, R. and {Barriere}, J. -C. and {Barros}, B.~J. and {Barthelemy}, A. and {Bartolo}, N. and {Basset}, A. and {Battaglia}, P. and {Battisti}, A.~J. and {Baugh}, C.~M. and {Baumont}, L. and {Bazzanini}, L. and {Beaulieu}, J. -P. and {Beckmann}, V. and {Belikov}, A.~N. and {Bel}, J. and {Bellagamba}, F. and {Bella}, M. and {Bellini}, E. and {Benabed}, K. and {Bender}, R. and {Benevento}, G. and {Bennett}, C.~L. and {Benson}, K. and {Bergamini}, P. and {Bermejo-Climent}, J.~R. and {Bernardeau}, F. and {Bertacca}, D. and {Berthe}, M. and {Berthier}, J. and {Bethermin}, M. and {Beutler}, F. and {Bevillon}, C. and {Bhargava}, S. and {Bhatawdekar}, R. and {Bianchi}, D. and {Bisigello}, L. and {Biviano}, A. and {Blake}, R.~P. and {Blanchard}, A. and {Blazek}, J. and {Blot}, L. and {Bosco}, A. and {Bodendorf}, C. and {Boenke}, T. and {B{\"o}hringer}, H. and {Boldrini}, P. and {Bolzonella}, M. and {Bonchi}, A. and {Bonici}, M. and {Bonino}, D. and {Bonino}, L. and {Bonvin}, C. and {Bon}, W. and {Booth}, J.~T. and {Borgani}, S. and {Borlaff}, A.~S. and {Borsato}, E. and {Bose}, B. and {Botticella}, M.~T. and {Boucaud}, A. and {Bouche}, F. and {Boucher}, J.~S. and {Boutigny}, D. and {Bouvard}, T. and {Bouwens}, R. and {Bouy}, H. and {Bowler}, R.~A.~A. and {Bozza}, V. and {Bozzo}, E. and {Branchini}, E. and {Brando}, G. and {Brau-Nogue}, S. and {Brekke}, P. and {Bremer}, M.~N. and {Brescia}, M. and {Breton}, M. -A. and {Brinchmann}, J. and {Brinckmann}, T. and {Brockley-Blatt}, C. and {Brodwin}, M. and {Brouard}, L. and {Brown}, M.~L. and {Bruton}, S. and {Bucko}, J. and {Buddelmeijer}, H. and {Buenadicha}, G. and {Buitrago}, F. and {Burger}, P. and {Burigana}, C. and {Busillo}, V. and {Busonero}, D. and {Cabanac}, R. and {Cabayol-Garcia}, L. and {Cagliari}, M.~S. and {Caillat}, A. and {Caillat}, L. and {Calabrese}, M. and {Calabro}, A. and {Calderone}, G. and {Calura}, F. and {Camacho Quevedo}, B. and {Camera}, S. and {Campos}, L. and {Ca{\~n}as-Herrera}, G. and {Candini}, G.~P. and {Cantiello}, M. and {Capobianco}, V. and {Cappellaro}, E. and {Cappelluti}, N. and {Cappi}, A. and {Caputi}, K.~I. and {Cara}, C. and {Carbone}, C. and {Cardone}, V.~F. and {Carella}, E. and {Carlberg}, R.~G. and {Carle}, M. and {Carminati}, L. and {Caro}, F. and {Carrasco}, J.~M. and {Carretero}, J. and {Carrilho}, P. and {Carron Duque}, J. and {Carry}, B.},
        title = "{Euclid: I. Overview of the Euclid mission}",
      journal = {A&A},
         year = 2025,
        month = may,
       volume = {697},
          eid = {A1},
        pages = {A1},
          doi = {10.1051/0004-6361/202450810},
archivePrefix = {arXiv},
       eprint = {2405.13491},
 primaryClass = {astro-ph.CO},
       adsurl = {https://ui.adsabs.harvard.edu/abs/2025A&A...697A...1E}
}

@article{CosmoSIS,
    author = "Zuntz, Joe and Paterno, Marc and Jennings, Elise and Rudd, Douglas and Manzotti, Alessandro and Dodelson, Scott and Bridle, Sarah and Sehrish, Saba and Kowalkowski, James",
    title = "{CosmoSIS: modular cosmological parameter estimation}",
    eprint = "1409.3409",
    archivePrefix = "arXiv",
    primaryClass = "astro-ph.CO",
    reportNumber = "FERMILAB-PUB-14-408-A",
    doi = "10.1016/j.ascom.2015.05.005",
    journal = "Astronomy and Computing",
    volume = "12",
    pages = "45--59",
    year = "2015"
}

@ARTICLE{Cobaya,
       author = {{Torrado}, Jes{\'u}s and {Lewis}, Antony},
        title = "{Cobaya: code for Bayesian analysis of hierarchical physical models}",
      journal = {Journal of Cosmology and Astroparticle Physics},
         year = 2021,
        month = may,
       volume = {2021},
       number = {5},
          eid = {057},
        pages = {057},
          doi = {10.1088/1475-7516/2021/05/057},
archivePrefix = {arXiv},
       eprint = {2005.05290},
 primaryClass = {astro-ph.IM},
       adsurl = {https://ui.adsabs.harvard.edu/abs/2021JCAP...05..057T}
}

@ARTICLE{EP-CLOE4,
       author = {{Euclid Collaboration: Martinelli}, M. and {Pezzotta}, A. and {Sciotti}, D. and others},
       title = "{Euclid preparation. Cosmology Likelihood for Observables in Euclid (CLOE). 4: Validation and Performance}",
      journal = {A\&A, accepted},
         year = 2025,
        month = oct,
          eid = {arXiv:2510.09141},
        pages = {arXiv:2510.09141},
          doi = {10.48550/arXiv.2510.09141},
archivePrefix = {arXiv},
       eprint = {2510.09141},
 primaryClass = {astro-ph.CO},
       adsurl = {https://ui.adsabs.harvard.edu/abs/2025arXiv251009141E}
}

@ARTICLE{EP-CLOE3,
       author = {{Euclid Collaboration: Ca{\~n}as-Herrera}, G. and {Goh}, L.~W.~K. and {Blot}, L. and others},
       title = "{Euclid preparation. Cosmology Likelihood for Observables in Euclid (CLOE). 3. Inference and Forecasts}",
      journal = {A\&A, accepted},
         year = 2025,
        month = oct,
          eid = {arXiv:2510.09153},
        pages = {arXiv:2510.09153},
          doi = {10.48550/arXiv.2510.09153},
archivePrefix = {arXiv},
       eprint = {2510.09153},
 primaryClass = {astro-ph.CO},
       adsurl = {https://ui.adsabs.harvard.edu/abs/2025arXiv251009153E}
}

@ARTICLE{EP-CLOE2,
       author = {{Euclid Collaboration: Joudaki}, S. and {Pettorino}, V. and {Blot}, L. and others},
       title = "{Euclid preparation. Cosmology Likelihood for Observables in Euclid (CLOE). 2. Code implementation}",
      journal = {A\&A, accepted},
         year = 2026,
        month = mar,
          eid = {arXiv:2603.22475},
        pages = {arXiv:2603.22475},
archivePrefix = {arXiv},
       eprint = {2603.22475},
 primaryClass = {astro-ph.CO},
       adsurl = {https://ui.adsabs.harvard.edu/abs/2026arXiv260322475E}
}

@ARTICLE{EP-CLOE1,
       author = {{Euclid Collaboration: Cardone}, V.~F. and {Joudaki}, S. and {Blot}, L. and others},
       title = "{Cosmology Likelihood for Observables in \textbackslashEuclid (CLOE). 1. Theoretical recipe}",
      journal = {A\&A, accepted},
         year = 2025,
        month = oct,
          eid = {arXiv:2510.09118},
        pages = {arXiv:2510.09118},
          doi = {10.48550/arXiv.2510.09118},
archivePrefix = {arXiv},
       eprint = {2510.09118},
 primaryClass = {astro-ph.CO},
       adsurl = {https://ui.adsabs.harvard.edu/abs/2025arXiv251009118E}
}

@article{CosmoLike,
    author = {Krause, Elisabeth and Eifler, Tim},
    title = {cosmolike – cosmological likelihood analyses for photometric galaxy surveys},
    journal = {Monthly Notices of the Royal Astronomical Society},
    volume = {470},
    number = {2},
    pages = {2100-2112},
    year = {2017},
    month = {05},
    issn = {0035-8711},
    doi = {10.1093/mnras/stx1261},
    url = {https://doi.org/10.1093/mnras/stx1261},
    eprint = {https://academic.oup.com/mnras/article-pdf/470/2/2100/18145458/stx1261.pdf},
}

\end{document}